\documentstyle[preprint,aps]{revtex}
\newcommand{\im}{\rightarrow}
\begin{document}
\draft
\title{ Phase Mixing of Nonlinear Plasma Oscillations in an Arbitrary
Mass Ratio Cold Plasma }
\author{Sudip Sen Gupta and Predhiman K.
Kaw } 
\address{Institute for Plasma Research, Bhat, Gandhinagar
382 428, India.}
\date{\today}
\maketitle

\begin{abstract}
	Nonlinear plasma oscillations in an arbitrary mass ratio cold plasma
have been studied using 1-D particle-in-cell simulation. In contrast to earlier
work for infinitely massive ion plasmas it has been found
that the oscillations phase mix away at any amplitude and that the 
rate at which phase mixing occurs, depends on
the mass ratio 
( $ \Delta = m_{-}/m_{+} $ ) and the amplitude.
A perturbation theoretic calculation carried upto third order predicts 
that the normalized phase mixing time $\omega_{p-} t_{mix}$
depends on the amplitude $A$ and the mass ratio $\Delta$ as $\sim
[(A^{2}/24)(\Delta/\sqrt{1 + \Delta})]^{-1/3}$.
We have confirmed this scaling in our simulations 
and conclude that stable non-linear oscillations which never
phase mix,
exist only for the ideal case with $ \Delta = 0.0 $ and $ A < 0.5 $.
These cold plasma results may have direct relevance to recent experiments
on superintense laser beam plasma interactions with applications to particle
acceleration, fast ignitor concept etc.
\end{abstract}

	The physics of the damping of nonlinear cold plasma oscillations
is a topic of considerable fundamental interest since it is the simplest
nonlinear collective irreversible phenomenon characterising the plasma state.  
It also has wide applications to a number of problems of current interest
such as particle acceleration by wakefields and beat waves created by
intense lasers or particle beams, the fast ignitor concept in
inertial fusion where relativistically intense coupled
electromagnetic - plasma waves modes propagate deep into overdense
plasmas to create a 'hot spark' and a number of other astrophysical /
laboratory / device based plasma experiments where intense plasma
oscillations are generated. 
The conventional thinking about the physics of
this interaction is well illustrated by the exact solution for nonlinear
one - dimensional cold plasma fluid equations with infinitely massive
ions. These exact solutions may be obtained by transforming to Lagrangian
coordinates as shown in \cite{daw,kun,david} or using stream functions \cite{kal}.
The exact solution shows that coherent oscillations
at the plasma frequency $\omega_{p}$ are maintained indefinitely over the
region of initial excitation, provided the normalized amplitude of the initial
density perturbation $A ( \equiv \frac{\delta n}{n} )$ is kept below $0.5$. 
For $A > 0.5$, one expects and observes wave breaking and fine scale mixing
of various parts of the oscillation \cite{daw}. Mathematically, the
electron number density blows up at $A=0.5$; this is because the Jacobian of
transformation from Eulerian to Lagrangian coordinates goes to zero as $A
\rightarrow 0.5$ and the transformation is no longer unique. Physically,
this is equivalent to crossing of electron trajectories which leads to
multistream motions and wave breaking as discussed in \cite{daw}. Studies
of wave breaking and phase mixing damping are based on  numerical
simulations.

The above description is adequate when the background positive species are
infinitely massive $( \frac{m_{-}}{m_{+}} \equiv \Delta \rightarrow 0
)$ and are uniformly distributed. 
If the background is inhomogeneous, then as was shown by
Dawson \cite{daw}, cold plasma oscillations phase mix away in a time scale
$t \sim \frac{\pi}{2 (d\omega_{p}/dx) X}$, at arbitrarily low amplitudes.
For a sinusoidal distribution of background species, such a phenomenon in 
the form of mode coupling of a long wavelength mode to short wavelength
modes was observed by Kaw et. al. \cite{kaw}. They found that, the time scale in which 
energy goes from long wavelength mode to short wavelength mode is 
$t \sim \frac{2}{\epsilon \omega_{p0}}$, where '$\epsilon$' is the amplitude
of the background inhomogeneity. The exact solution for the cold plasma oscillations
in a fixed sinusoidal background was given by Infeld et. al. \cite{inf}
who described phase mixing in terms of electron density burst.

In this paper we show that the phenomenon of phase mixing will also
occur in a homogeneous plasma at arbitrarily low amplitudes, provided
the background 
positive species are allowed to move $(\Delta \neq 0)$. This is because
the background species respond to ponderomotive forces either directly
or through low frequency self-consistent fields and thereby acquire
inhomogenities in space.
Such an effect has been observed in electron
positron plasmas $( \Delta = 1 )$ by Stewart \cite{ste}. In plasmas with
finite temperature, it is well known that plasma waves dig cavities by
ponderomotive forces and get trapped in them; this is the physics of strong
turbulence of Langmuir waves as elucidated by Zakharov \cite{zak} and leads
to envelope soliton formation in one - dimension and collapse phenomenon in
2-D and 3-D. In a cold plasma, stationary states cannot form even in 1-D
because there is no thermal pressure effect to counterbalance the
ponderomotive forces. The result is that the density cavities being dug by
plasma oscillations have an amplitude which increases secularly in time
. Similarly, the response of plasma oscillations to the presence
of density cavities is also different from that of Zakharov problem. In the
Zakharov problem, the thermally dispersive plasma waves get trapped in
density cavities forming localized wavepackets. Here, the inhomogeneity of
the cold plasma ( due to the self - consistently generated
perturbation ) causes different parts of the plasma oscillation to
oscillate at different frequencies \cite{daw} \cite{kaw} \cite{inf}
resulting in intense phase mixing of plasma oscillations.
Thus we physically
expect that if the background species is allowed to move and get
redistributed into inhomogeneous clumps of density, the phase mixing damping of cold plasma oscillations should
come in at any amplitude and is not restricted to waves with $A > 0.5$. 
It may be emphasised here that for many applications involving interaction
of superintense laser beams with plasmas ( such as particle accleration by 
wake fields, penetration into overdense plasmas etc. ) the cold plasma
limit considered by us is more relevant than the Zakharov description, 
because typically the plasma wave intensities are such that $\mid E \mid^{2}/4\pi nT \gg 1$.

In this paper we carry out particle simulations for elucidating the physics
of phase mixing damping of nonlinear cold plasma oscillations in an
arbitrary mass ratio plasma ($\Delta$ arbitrary). 
We also present a perturbation - theoretic analysis to give a quantitative
estimate of the phase mixing time for moderate amplitude oscillations and
compare it with simulation.

We start with the cold plasma equations {\it viz.}
the continuity equations and the equations of motion for the two species
and the Poisson equation. 
We introduce new variables $V, v, \delta n_{d}$
and $\delta n_{s}$ defined as $V = v_{+} + v_{-}$, $v = v_{+} - v_{-}$,
$\delta n_{d} = \delta n_{+} - \delta n_{-} = n_{+} - n_{-}$ and
$\delta n_{s} = \delta n_{+} + \delta n_{-} = n_{+} + n_{-} - 2$ to
write the cold plasma equations in the form

\begin{eqnarray}
\partial_{t} \delta n_{d} + \partial_{x} [ v + \frac{V \delta n_{d} + v
\delta n_{s}}{2} ] = 0 \\
\partial_{t} \delta n_{s} + \partial_{x} [ V + \frac{V \delta n_{s} + v
\delta n_{d}}{2} ] = 0 \\
\partial_{t} V + \partial_{x} ( \frac{ V^{2} + v^{2}}{4} ) = -(1 - \Delta)
E \\
\partial_{t} v + \partial_{x} ( \frac{Vv}{2} ) = (1 + \Delta) E \\
\partial_{x} E = \delta n_{d}
\end{eqnarray}

Note that we have used the normalizations: $n_{\pm} \im n_{\pm}/n_{0}$,
$x \im kx$, $t \im \omega_{p-}t$, $v_{\pm} \im v_{\pm}/\omega_{p-} k^{-1}$,
$E \im E/(4 \pi n_{0} e k^{-1} )$, with $\omega_{p-}^{2} = 4 \pi n_{0} e^{2} / m_{-}$
and $\Delta = m_{-}/m_{+}$. 

Using $n_{-}(x,0) = 1 + \delta \cos kx$, $n_{+}(x,0) = 1$ and $v_{\pm}(x,0)
= 0$, as initial conditions, the solutions of the linearised equations are

\begin{eqnarray}
\delta n_{d}^{(1)} = A \cos kx \cos \omega_{p} t \\
E^{(1)} = \frac{A}{k} \sin kx \cos \omega_{p} t \\
\delta n_{s}^{(1)} = \frac{1 - \Delta}{1 + \Delta} A \cos kx ( 1 - \cos
\omega_{p} t ) - A \cos kx \\
V^{(1)} = - \frac{1 - \Delta}{k \omega_{p}} A \sin kx \sin \omega_{p} t \\
v^{(1)} = \frac{1 + \Delta}{k \omega_{p}} A \sin kx \sin \omega_{p} t  
\end{eqnarray}

where $A = - \delta$ and $\omega_{p}^{2} = 1 + \Delta$.
At this level of approximation, the solutions show coherent oscillations at
the plasma
frequency $\omega_{p}$. Both the species oscillate with the same frequency
which is independent of position.

In the 2nd order, the solutions are expressed as:
\begin{eqnarray}
\delta n_{d}^{(2)} = - A^{2} \cos 2kx [ \frac{1 - \Delta}{1 + \Delta}
( \frac{1}{2} + 
\frac{1}{4} \omega_{p}t \sin \omega_{p} t + \frac{1}{2} \cos 2 \omega_{p} t
- \cos \omega_{p} t ) - \frac{1}{4} \omega_{p}t \sin \omega_{p}t ] \\
\delta E^{(2)} = - \frac{A^{2}}{2k} \sin 2kx [ \frac{1 - \Delta}{1 +
\Delta}
( \frac{1}{2} + 
\frac{1}{4} \omega_{p}t \sin \omega_{p} t + \frac{1}{2} \cos 2 \omega_{p} t
- \cos \omega_{p} t ) - \frac{1}{4} \omega_{p}t \sin \omega_{p}t ]
\end{eqnarray}
\begin{eqnarray}
\delta n_{s}^{(2)} = \frac{A^{2}}{2} \cos 2kx [ \frac{\Delta}{1 + \Delta}
t^{2} - \frac{\Delta (1 - \Delta)}{(1 + \Delta)^{2}} \omega_{p}t \sin
\omega_{p}t - \frac{3}{8}(1 - \cos 2 \omega_{p}t) - \\
(\frac{1 - \Delta}{1 + \Delta})^{2}(2 \cos \omega_{p}t -
\frac{5}{8} \cos 2
\omega_{p}t - \frac{11}{8}) ] \nonumber 
\end{eqnarray}
\begin{eqnarray}
V^{(2)} = - \frac{A^{2}}{2 k} \sin 2kx [\frac{\Delta}{1 + \Delta} t 
+ \frac{\omega_{p}}{2}(\frac{1 - \Delta}{1 + \Delta})^{2} ( \frac{1 - 3
\Delta}{2(1 - \Delta)}\omega_{p} t
\cos \omega_{p} t + \\
\frac{7 - 5 \Delta}{2(1 - \Delta)} \sin  \omega_{p} t - \frac{5}{4} \sin 2
\omega_{p} t) - 
\frac{1}{8} \omega_{p} \sin 2 \omega_{p} t ] \nonumber
\end{eqnarray}
\begin{eqnarray}
v^{(2)} = - \frac{A^{2} \omega_{p}}{8k} \sin 2kx [ 
\sin \omega_{p} t - \omega_{p} t \cos \omega_{p} t - \frac{1 - \Delta}{1 +
\Delta} ( 2 \sin 2 \omega_{p} t - 3 \sin \omega_{p} t - \omega_{p} t \cos
\omega_{p} t) ]
\end{eqnarray}

The 2nd order solutions clearly exhibit generation of 2nd harmonic in
space and time as well as bunching of plasma particles in space.
Both of
these features are also evident in the solution of Kaw et. al. \cite{kaw}
and Infeld et. al. \cite{inf}; but in contrast to their work, where the
background ion density was kept fixed in time, here the density of the
plasma particles self - consistently changes with time as $\sim
t^{2}$, as seen in the expression for $\delta n_{s}^{(2)}$. 
Because of variation of plasma density with time, the phase mixing
of an initial coherent oscillation happens much faster in this case. To 
make an estimate of the phase mixing time, consider the charge
density equation ( $\delta n_{d}$ in this case ). The equation for $\delta
n_{d}$ correct upto third order stands as

\begin{eqnarray}
\partial_{tt} \delta n_{d} + \omega_{p}^{2}[ 1 + \frac{1}{2} ( \delta n_{s}^{(1)}
+ \delta n_{s}^{(2)}) ] \delta n_{d} \approx 0
\end{eqnarray}

In the above equation, if we neglect the 2nd order term, then we
essentially get the same phase mixing time as in Ref. \cite{kaw} modified
by a factor which depends on $\Delta$. Now
taking only the leading order secular terms from the expressions of $\delta
n_{s}^{(1)}$
and $\delta n_{s}^{(2)}$ ( there are no secular terms in $\delta n_{s}^{(1)}$ )
we get 
\begin{eqnarray}
\partial_{tt} \delta n_{d} + \omega_{p}^{2}[ 1 + \frac{A^{2}t^{2} \Delta}{4 \omega_{p}^{2}}
\cos 2 k x ] \delta n_{d}\approx 0
\end{eqnarray}

 Using the initial conditions $\delta n_{d} = A \cos kx$ and $\partial_t
 \delta n_{d} = 0$ the WKB solution of the above equation is

\begin{eqnarray}
\delta n_{d} \approx A \cos kx \sum_{n = - \infty}^{n = \infty}
\cos( \omega_{p} t + \frac{n \pi}{2} - 2 n k x ) J_{n}( \frac{A^{2}t^{3} \Delta}{24 \sqrt{1 + \Delta}} )
\end{eqnarray}

 The above expression clearly shows that the energy which was initially in
the primary wave at mode $k$ goes into higher and higher harmonics as time
progresses. This can be interpreted as damping of the primary wave due to
mode coupling to higher and higher modes. Microscopically, as the
plasma particles oscillate at the local plasma frequency, they gradually go
out of phase and eventually the initial coherence is lost. Because of
generation of higher and higher harmonics with time, the charge density
becomes more and more spiky and as a result the electric field gradients
become more and more steep. This does not go on indefinitely. In reality,
the density peaks get limited by thermal effects with the Landau damping of
high $k$ modes by resonant particles coming into picture. This process
takes energy from the high $k$ modes and puts it on the particles, thereby
raising their temperature, which in turn limits the density peaks by
exerting a pressure gradient. 
The time scale in which the initial coherence is lost ( or the phase
mixing time ) can be seen 
from equation (18) as
$\omega_{p-} t_{mix}$ scale as $\sim [ A^{2} \Delta / (24 \sqrt{1 +
\Delta}) ]^{- 1/3}$. It  shows that  only for the ideal case $\Delta = 0.0$
(infinitely massive ions), phase mixing time is infinity, i.e. the initial
coherence is maintained indefinitely \cite{kun,david}. For an actual electron
- ion plasma, $\Delta$ although small, is finite and hence plasma
oscillations in it phase mix away at arbitrarily small amplitudes and 
in a time scale dictated by the amplitude
of the initial perturbation.

Now we present results from a 1-D particle-in-cell simulation which confirms
our scaling of phase mixing time. For numerical simulation, we have used a
1-dimensional model with periodic boundary conditions and have followed
5120 electrons and as many positively charged particles ( the plasma taken
as a whole is neutral ) in their own self - consistent fields. The
particles are initially at rest and the system is set into motion by giving
a density perturbation of the form $n_{-} = 1 + \delta \cos kx$ to the
electrons. In the simulation, we follow the time development of various
modes of charge density ($\delta n_{d}$). To compare with our theoretical
model, we rewrite equation (18) as
\begin{eqnarray}
\delta n_{d} = \frac{A}{2} \sum_{n=- \infty}^{n=\infty} J_n[\alpha(t)] [ \cos(
\omega_{p}t + \frac{n \pi}{2}) \{ \cos(2n + 1)kx + \cos(2n - 1)kx \} + \\
\sin(\omega_{p}t + \frac{n \pi}{2}) \{ \sin(2n+1)kx + \sin(2n - 1)kx \} ]
\nonumber 
\end{eqnarray}

where $\alpha(t) = \frac{A^{2}t^{3}}{24} \frac{\Delta}{\sqrt{1 +
\Delta}}$. The amplitude of the first fourier mode can be seen from the
above equation as

\begin{eqnarray}
\mid \delta n_{d} \mid_{n = 1} = \frac{A}{2} [ J_{0}^{2}(\alpha(t)) +
J_{1}^{2}( \alpha(t)) ]^{\frac{1}{2}}
\end{eqnarray}

It is clear from equation (19), that upto the order of approximation
considered, there are no even number modes in the system. Fig. (1)
show temporal variation of $\mid \delta n_{d} \mid_{n=1}$ for $\Delta = 1.0$
and $A = 0.05$. The dotted curve is the simulation result and the solid 
line shows our expression (20) for the envelope of the oscillations.
It is clear from the figure that our approximate expression (20)
captures the early evolution of the plasma quite well.
Fig. (2) shows the variation of $\tau_{mix} = \omega_{p} t_{mix}$ with $\Delta$ for
a fixed $A = 0.1$ (curve(1)) and with $A$ for a fixed $\Delta=0.01$
(curve(2)). These curves clearly confirm our formula for phase mixing
time.

%

In conclusion, we have demonstrated that nonlinear plasma oscillations
in a cold homogeneous plasma, phase mix away at 
arbitrarily low amplitudes. This
is because during the course of motion the plasma particles respond to
ponderomotive forces, acquiring inhomogeneity and thereby
making the plasma frequency a function of
space. As a result, electrons at different locations oscillate with different
( local ) plasma frequencies and the imposed plasma wave losses coherence.
The formation of density clumps can also be seen
from the Zakharov equations \cite{zak} for a warm electron - ion plasma. 
According to Zakharov, the slow
variation ( in the ion time scale ) of the background density in the 
presence of a high frequency oscillation is governed by
$\partial_{tt} \delta n_{s} - T \partial_{xx}
\delta n _{s} = \partial_{xx} \mid E \mid^{2}$. In the limit when the
thermal term balances the ponderomotive force term ( i.e. $\frac{\delta
n_{s}}{n_{0}} \approx - \frac{\mid E \mid^{2}}{T}$ ), we get caviton
solutions in 1-D which are unstable to transverse
perturbations. In the other limit, when $\frac{\mid E \mid^{2}}{T} \gg 1$,
it is the $\partial_{tt} \delta n_{s}$ term which dominates, 
and the Zakharov equation shows density rising as $\sim t^{2}$.
This is the same scaling as obtained by us using a perturbative approach.
The density inhomogenities thus created lead to phase mixing 
and collapse of cavitons.
From this we infer that a cold 1-D plasma exhibits
a 'Langmuir collapse' phenomenon similar to what is seen in a warm plasma
in 2 or 3 dimensions. The time scale of collapse is of the order 
$\sim (\frac{A^{2}}{24}\frac{\Delta}{\sqrt{1+\Delta}})^{-\frac{1}{3}}$
plasma periods. Recent experiments on plasma acceleration by laser 
wakefields have shown \cite{joshi} that wave breaking of excited plasma oscillations
plays a major role in the final acceleration process ; similar physics is
likely to be important in the fast ignitor concept of laser fusion. 
We expect the processes discussed in the present paper 
to play some role in such experiments. It should be noted that the simulation
results presented in this letter are non - relativistic. For many 
experimental situations the jitter velocity of electrons is relativistic and 
we expect the mass ratio $\Delta$ to be replaced by $\Delta_{eff} \approx m_{eff}/m_{i}$
where $m_{eff}/m_{i} \gg 1$. Under these conditions, the phase mixing effects
considered by us should become more important. Such investigations are in
progress and will be presented elsewhere.

\newpage
{\bf Figure Captions}\\[0.2in]
Fig. 1: $\mid \delta n_{d} \mid_{n=1}$ vs. $t/T_{P}$ for $\Delta = 1.0$ and $A = 0.05$\\
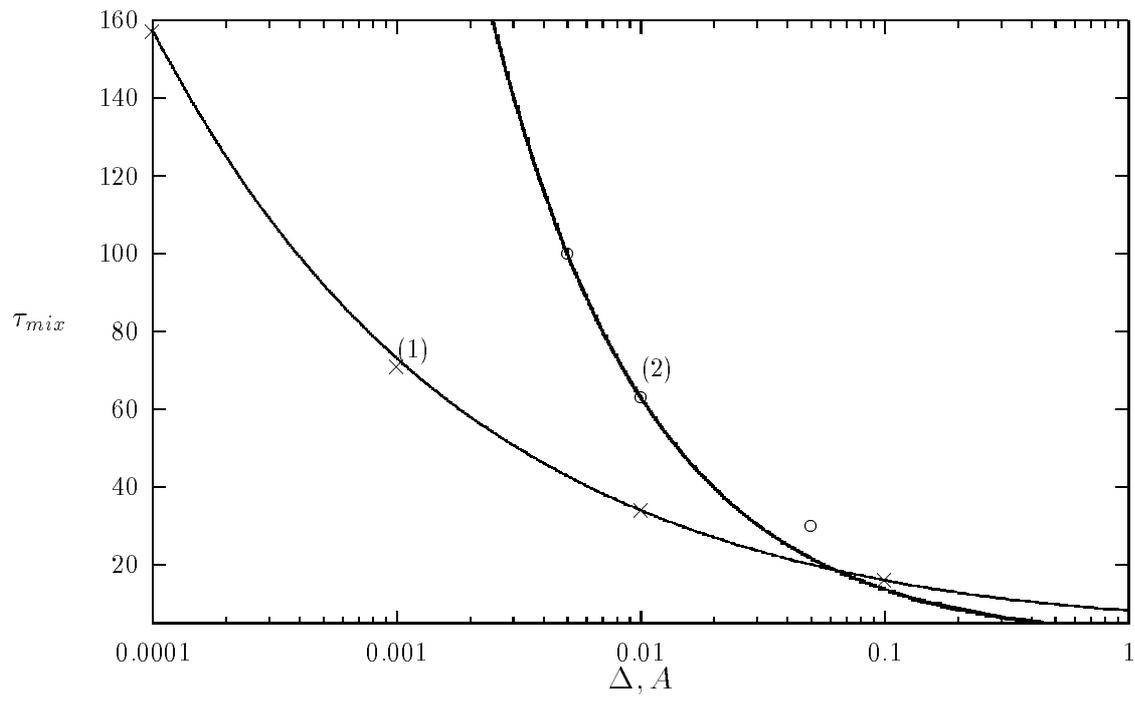
Fig. 2: $\tau_{mix}$ vs. $\Delta$ and $A$
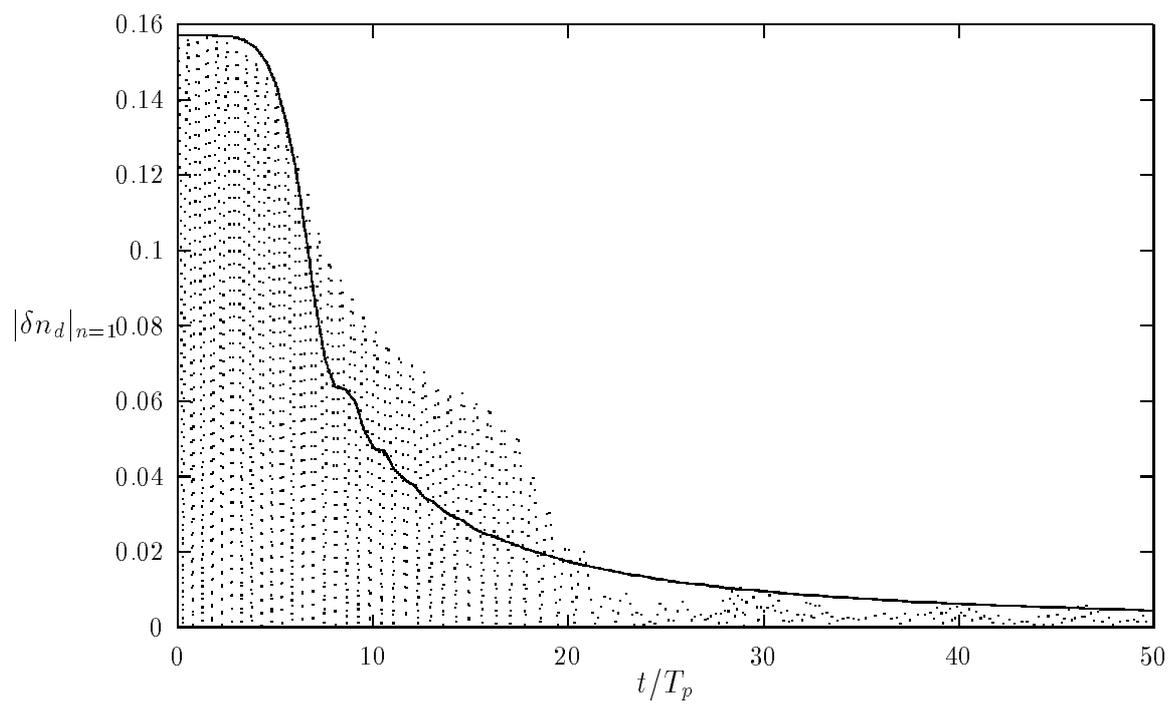
\end{document}